\documentclass[pra,twocolumn,letterpaper,accepted=2018-03-23]{quantumarticle}

\usepackage[margin=1in]{geometry}
\usepackage[english]{babel}
\usepackage{ucs}
\usepackage[utf8x]{inputenc}
\usepackage{amsmath}
\usepackage{amssymb}
\usepackage{amsthm}
\usepackage{graphicx}
\usepackage{bm}
\usepackage{bbm}
\usepackage{braket}
\usepackage{empheq}
\usepackage{subcaption}
\usepackage[colorlinks=true]{hyperref}
\usepackage[font=footnotesize,labelfont=bf]{caption}

\graphicspath{{Images/}{../Images/}}

\newcommand{\hrho}{\hat{\rho}}
\newcommand{\balpha}{\overline{\alpha}}

\begin{document}

\title{Dissipative stabilization of entangled cat states using a driven Bose-Hubbard dimer}

\author{M. Mamaev}
\affiliation{Department of Physics, McGill University,  Montr\'eal, Qu\'ebec, Canada.}

\author{L. C. G. Govia}
\affiliation{Institute for Molecular Engineering, University of Chicago, 5640 S. Ellis Ave., Chicago, IL 60637}

\author{A. A. Clerk}
\affiliation{Institute for Molecular Engineering, University of Chicago, 5640 S. Ellis Ave., Chicago, IL 60637}

\date{}

\begin{abstract}
We analyze a modified Bose-Hubbard model, where two cavities having on-site Kerr interactions are subject to two-photon driving and correlated dissipation.
We derive an exact solution for the steady state of this interacting driven-dissipative system, and use it show that the system permits the preparation and stabilization of pure entangled non-Gaussian states, so-called entangled cat states.  Unlike previous proposals for dissipative stabilization of such states, our approach requires only a linear coupling to a single engineered reservoir (as opposed to nonlinear couplings to two or more reservoirs).  Our scheme is within the reach of state-of-the-art experiments in circuit QED.
\end{abstract}

\maketitle

\section{Introduction}
To harness the full potential of continuous variable quantum computing \cite{RevModPhys.77.513}, it is necessary to prepare and ideally stabilize non-trivial entangled states of multiple bosonic modes.  An ideal approach here is reservoir engineering \cite{poyatos1996reservoirEngineering},
where controlled interactions with tailored dissipation are used to accomplish both these goals.
The simplest kind of entangled bosonic states are the two-mode squeezed states, and there exist many closely-related reservoir engineering protocols which allow their stabilization (see e.g.~\cite{Parkins2006,Krauter2011,Muschik2011,woolley2014reservoirEngineering}).

While useful in certain applications, two-mode squeezed states are Gaussian, and hence describable by a positive-definite Wigner function.  The ability to generate and stabilize truly non-Gaussian entangled states (which exhibit negativity in their Wigner functions) would provide an even more powerful potential resource.  A canonical example of such states are two-mode entangled Schr\"odinger cat states \cite{Haroche2006}.  These states are integral to bosonic codes for quantum computing \cite{vlastakis2013cats,mirrahimi2014yaleCats,wang2016yaleCats}, and are equivalent to so-called ``entangled coherent states", which have applications in many aspects of quantum information \cite{sanders2012coherentStatesReview}.

Dissipative stabilization of non-Gaussian states is in principle possible, but requires more complicated setups than in the Gaussian case; in particular, nonlinearity is now an essential resource.  This complexity is present even when trying to reservoir engineer cat states in a single mode:  for example, a recently implemented protocol requires one to engineer a two-photon loss operator \cite{mirrahimi2014yaleCats,Leghtas2015}. There exist protocols for stabilizing entangled cat states \cite{Sarlette2012,arenz2013entangledCatGeneration} based on passing a stream of few-level atoms through a pair of cavities.  These approaches effectively make use of either a complex nonlinear dissipator or highly nonlinear cavity driving.

\begin{figure}[t]
\centering
\includegraphics[width=0.9 \linewidth]{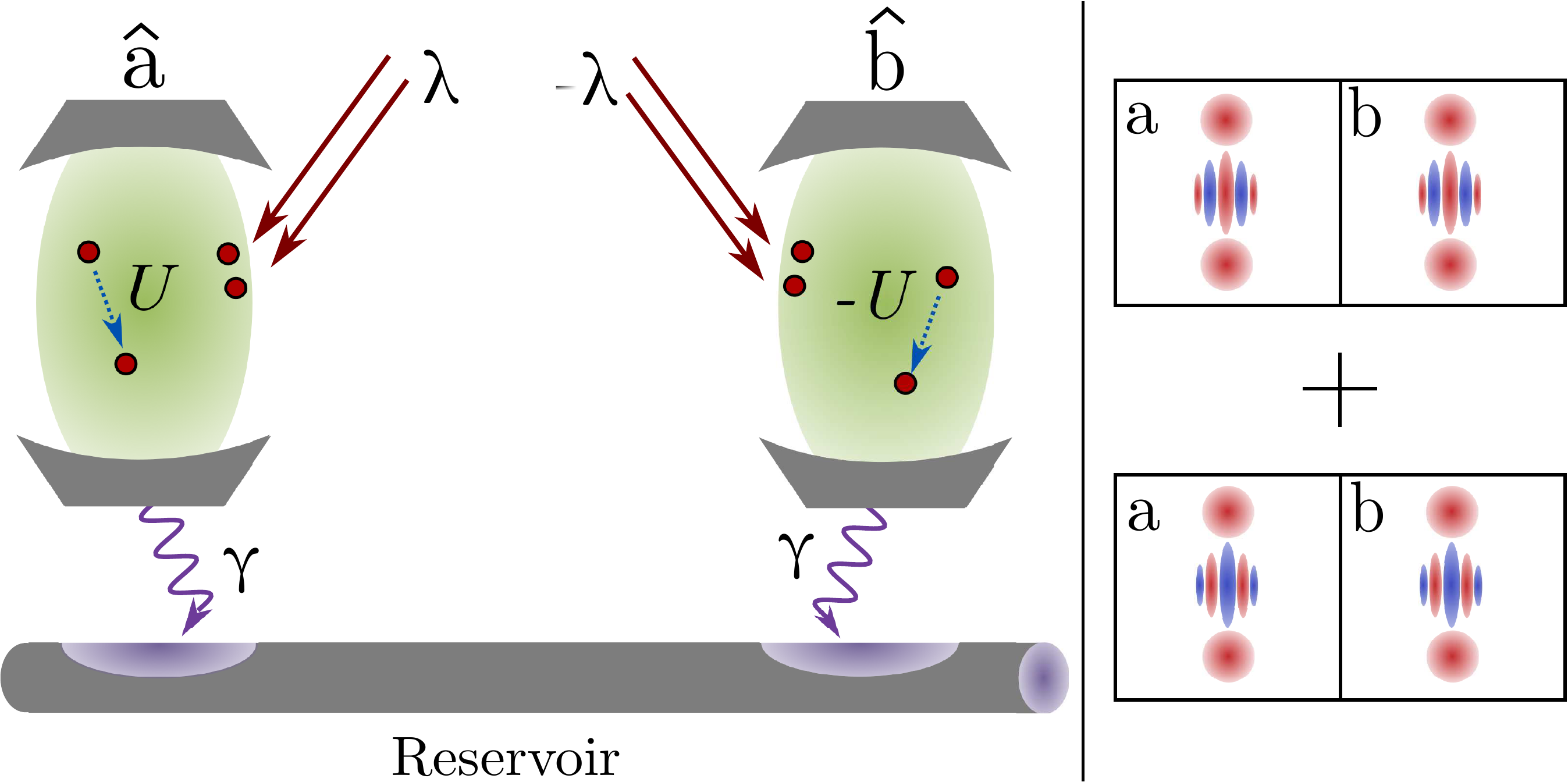}
\caption{Left panel:   Schematic of the system:  two bosonic modes $\hat{a}$, $\hat{b}$ are subjected to two-photon driving (amplitude $\lambda$), and have on-site Kerr-interactions which are repulsive (attractive) for mode $\hat{a}$ (mode $\hat{b}$).  The two modes couple to a single engineered Markovian reservoir (e.g.~a waveguide) which mediates the dissipative equivalent of a tunneling interaction.  For appropriately matched parameters (see main text), the system has a unique pure steady state corresponding to an entangled Schr\"{o}dinger cat state.  This state is an equal superposition of each cavity being in an even-parity cat state, and each cavity being in an odd-parity cat state.  This is depicted schematically in the right panel, using Wigner functions for the cat states involved in the superpositions.}
\label{fig_schematic}
\end{figure}

In this paper, we present a surprisingly simple scheme for stabilizing a two-mode entangled cat-state.  Our setup is sketched in Fig.~\ref{fig_schematic}, and resembles the well-studied dissipative Bose-Hubbard dimer model (see e.g.~\cite{Pudlik2013,Grujic:2013aa,Hafezi2016,Casteels2016,Casteels2017,Noh:2017aa}), where two driven bosonic modes with on-site Kerr interactions are coupled via coherent tunneling.  We make some crucial modifications to this standard setup.  In our system, both cavities are subject to coherent two-photon (i.e.~parametric) driving, as opposed to the more standard linear driving.  Further, there is no direct tunneling between the cavities.  Instead, both cavities couple linearly to a common dissipative environment (the engineered reservoir), which mediates the dissipative equivalent of a tunnel interaction.  This dissipative coupling could be generated in many different ways, including by simply coupling both cavities passively to a low-loss waveguide \cite{Chang2012,Metelmann2015}.  Finally, we require one cavity to have an attractive interaction, while the other has a repulsive interaction.

Despite being an interacting open system, we are able to analytically solve for the stationary state, something that is not possible for the standard driven-dissipative Bose Hubbard dimer \cite{Hafezi2016}. Using our exact solution, we identify a simple parameter regime where the steady state is a pure entangled cat state.  Our scheme is also easily modified so that it stabilizes a two-dimensional manifold spanned by two even-parity cat states; this allows a potential two-cavity generalization of the cat-state encoded qubit introduced in Ref.~\cite{mirrahimi2014yaleCats}.  We stress that all the key ingredients required for our scheme are available in state of the art circuit QED setups (namely parametric drives, see e.g.~\cite{Yamamoto:2008aa,Krantz:2016aa,Mutus:2014aa}, strong on-site Kerr interactions \cite{kirchmair2012kerr} and engineered dissipation \cite{Sliwa:2015aa,Lecocq:2017aa,Lecocq:2016aa}).  While the relaxation rate associated with our stabilization scheme can become very slow for large photon number entangled cat states, we show explicitly that our scheme is still effective in state-of-the-art systems and resilient to low levels of internal cavity loss.

\section{Single-mode cat state stabilization}

\subsection{Dissipative stabilization of a single-mode squeezed state}

We introduce our scheme by first reviewing the well-understood reservoir engineering protocol for stabilizing a single-mode squeezed state
\cite{Cirac1993,Rabl2004,Kronwald2013}.  This protocol was recently realized experimentally both in optomechanics \cite{Schwab2015,Teufel2015,Sillanpaa2015} as well as in a trapped ion  system \cite{Kienzler2015}. The protocol is based on the observation that a squeezed state is the vacuum of a bosonic Bogoliubov operator  $\hat{\beta}[r] = \cosh(r)\hat{d} + \sinh(r)\hat{d}^\dagger$, where $\hat{d}$ is the annihilation operator for the mode of interest.  Hence, by simply cooling the mode $\hat{\beta}[r]$, one can stabilize a squeezed state.

The required dissipative cooling dynamics is obtained most simply from a Lindblad master equation for the reduced density matrix  $\hrho$ of the mode:
\begin{equation}
	\label{eq:BasicMEq}
	\frac{d}{dt}\hrho = \Gamma \left( \hat{z} \hrho \hat{z}^{\dagger} - \frac{1}{2}\{\hat{z}^{\dagger}\hat{z}, \hrho\} \right)
	\equiv \Gamma \mathcal{L}[\hat{z}] \hrho,
\end{equation}
with a jump operator $\hat{z}$ given by
\begin{equation}
	\label{eq:JumpOpSqueezed}
	\hat{z} \equiv  \mu \hat{d} + \nu \hat{d}^\dagger.
\end{equation}
$\Gamma$ represents the rate of the dissipative interaction (i.e.~the cooling rate), and the parameters $\mu, \nu$ are set to
\begin{equation}
	\mu = \cosh r, \,\,\, \nu = \sinh r,
\end{equation}
 making the jump operator $\hat{z}$ equal to the Bogoliubov-mode annihilation operator $\hat{\beta}[r]$.
 The unique steady state of this master equation is then the dark state of the jump operator $\hat{z}$, which is nothing else than the vacuum squeezed state with squeeze parameter $r$.

\subsection{Dissipative stabilization of a single-mode cat state}

To modify this scheme to stabilize a non-Gaussian state, we now introduce nonlinearity into the jump operator $\hat{z}$.
One simple way to do this is to make the parameter $\mu$ a photon-number dependent operator.  We take $\mu = \mu_1 \hat{d}^{\dagger}\hat{d}$, and consider a master equation of the form in Eq.~(\ref{eq:BasicMEq}) with jump operator
\begin{equation}
		\label{eq:JumpOpCat}
	\hat{z} =  \left( \mu_1 \hat{d}^{\dagger}\hat{d} \right) \hat{d} + \nu \hat{d}^{\dagger}.
\end{equation}
The steady states of this master equation correspond to the dark states of the jump operator $\hat{z}$.  Despite the nonlinearity, these dark states are easily found:  they correspond to a two-dimensional subspace spanned by even and odd Schr\"odinger cat states.  Defining even/odd cat-states in terms of coherent states $\ket{\alpha}$ as
\begin{equation}
	\Ket{\mathcal{C}_{\pm}(\alpha)}=\frac{\ket{\alpha}\pm\ket{-\alpha}}{\sqrt{2(1\pm e^{-2|\alpha|^2})}},
\end{equation}
we find that our jump operator has dark states
\begin{equation}
	\label{eq_nonlinearDarkStateCondition}
	\begin{aligned}
	\hat{z} \ket{\mathcal{C}_{\pm}(\sqrt{2}\balpha)}=0,
\end{aligned}
\end{equation}
where the amplitude $\balpha$ is given by
\begin{equation}
	\balpha  =i \sqrt{\nu / (2 \mu_1)}.
	\label{eq:alpha}
\end{equation}

For some applications, the ability to stabilize a manifold of dark-states can be extremely useful, see e.g.~\cite{mirrahimi2014yaleCats}.  However, for the simplest state preparation applications, it is ideal to have a unique steady state.  We would thus like to break the symmetry between the even and odd cat-states above, and stabilize, e.g., only the even parity cat.  To do this, we will simply interpolate between the jump operator in Eq.~(\ref{eq:JumpOpSqueezed}) (which stabilizes an even-parity squeezed state) and the nonlinear jump operator in
Eq.~(\ref{eq:JumpOpCat}).  We thus consider a modified jump operator
\begin{equation}
	\hat{z} = \left(\mu_0 + \mu_1 \hat{d}^{\dagger}\hat{d}\right) \hat{d}+ \nu \hat{d}^{\dagger} \equiv \hat{z}_{\rm cat}.
	\label{eq:zcat}
\end{equation}
As we show explicitly in Appendix \ref{app:Exact}, for $\mu_0 \neq 0$, the only possible dark states of this dissipator are even-parity states.  We thus find a {\it unique} dark state (and hence pure steady state of Eq.~(\ref{eq:BasicMEq})):
\begin{equation}
	\hat{z}_{\rm cat} \Ket{\tilde{C}[\mu_0,\mu_1,\nu]} = 0.
	 \label{eq:OneModeSS}
\end{equation}
One can obtain an analytic expression for this state, we discuss this further in Sec.~\ref{sec_exactSolution} (c.f. Eq.~(\ref{eq_exactSteadyState})). As could already be expected from the results above, we find that in the limit of small  $\mu_0$, this state asymptotically approaches an even-parity cat state:
\begin{equation}
	\lim_{\mu_0 / \mu_1 \rightarrow 0}
		\left| \Braket{ \tilde{C}[\mu_0,\mu_1,\nu] | \mathcal{C}_+[\sqrt{2} \balpha]} \right| \rightarrow 1.
\end{equation}
Even for small but finite
$\mu_0 / \mu_1$, $\ket{\tilde{C}[\mu_0,\mu_1,\nu]}$ has an extremely high fidelity with the ideal even parity cat-state (c.f.~Fig.~\ref{fig_squeezedToCat}).
We have thus shown how (in principle) the usual reservoir engineering recipe for stabilizing a single-mode squeezed state can be extended to stabilizing an even-parity
cat state.

\subsection{Auxiliary cavity method for realizing a nonlinear dissipator}

We now consider how one might realize the nonlinear jump operator in Eq.~(\ref{eq:zcat}).  The usual recipe is to introduce a second highly damped mode $\hat{c}$ which couples nonlinearly to the principle mode $\hat{d}$ of interest.  To that end, we consider a two-mode system whose non-dissipative dynamics is described by the Hamiltonian
\begin{align}
	\hat{H}_1 & =
		\Lambda \left( \hat{c}^\dagger \hat{z}_{\rm cat} + \hat{z}_{\rm cat}^\dagger \hat{c} \right) \nonumber \\
			& =
		\Lambda \hat{c}^{\dagger}\left[\mu_0 \hat{d} + \mu_1 (\hat{d}^{\dagger}\hat{d})\hat{d}
			+ \nu \hat{d}^{\dagger} \right] +h.c. \label{eq:H1}
\end{align}
The quadratic terms in this Hamiltonian correspond to a simple beam-splitter interaction ($\propto \mu_0$) and two-mode squeezing interaction ($\propto \nu$).  The remaining nonlinear term ($\propto \mu_1$) describes photon-number dependent tunneling.

We now add damping of the $\hat{c}$ mode (loss at rate $\gamma$), such that the full evolution of the two modes is given by
\begin{equation}
	\frac{d}{dt} \hrho =-i[\hat{H_1},\hrho]+\gamma\mathcal{L}[\hat{c}] \hrho.
	\label{eq:TwoModeMEq}
\end{equation}
In the limit where the $\hat{c}$ mode is heavily damped (i.e.~$\gamma \gg \Lambda$), standard adiabatic elimination yields a master equation of the form in
Eq.~(\ref{eq:BasicMEq}) with jump operator $\hat{z}_{\rm cat}$ and with $\Gamma \propto \Lambda^2 / \gamma$.  In this limit, our two-mode system has a unique steady state
analogous to that in Eq.~(\ref{eq:OneModeSS}):
\begin{align}
	\ket{\psi_{\rm ss}} & = \ket{0}_c \ket{\tilde{C}[\mu_0,\mu_1,\nu]}_d.
	\label{eq:TwoModeSS}
\end{align}
As discussed earlier, the state $\ket{\tilde{C}[\mu_0,\mu_1,\nu]}_d$ asymptotically approaches an even cat-state in the limit $\mu_0 / \mu_1 \rightarrow 0$.

We next make a crucial observation:  even if one is far from the large-damping limit, the above state is still the steady state of the full master equation in Eq.~(\ref{eq:TwoModeMEq}).  This follows directly from the fact that $\hat{H}_1$ and $\hat{c}$ both annihilate this state.  Thus, despite the nonlinearity in our two-mode system, we have found its steady state (analytically, c.f. Eq.~(\ref{eq_exactSteadyState})) for all values of the parameters $\gamma, \Lambda, \mu_0, \mu_1$ and $\nu$.


\section{Stabilization of two-mode entangled cat states}

The previous section described a setup for stabilizing a single-mode cat state in a cavity, using a nonlinear interaction with a damped auxiliary cavity.  We will now connect this result
to the stated goal of this paper:  stabilization of non-Gaussian entangled states using a system which resembles a driven-dissipative Bose-Hubbard dimer model.  Surprisingly, we will find that the resulting setup is much simpler than that described in Eq.~(\ref{eq:H1}).

We again start with our two modes $\hat{c}$ and $\hat{d}$, with the dissipative dynamics described in the previous section.  We now imagine that these modes are delocalized modes, describing equal superpositions of two distinct localized cavity modes $\hat{a}$ and $\hat{b}$:
\begin{equation}
	\label{eq_conversionABtoCD}
	\hat{a}=\frac{1}{\sqrt{2}}(\hat{c}+\hat{d}),\>\>\hat{b}=\frac{1}{\sqrt{2}}(\hat{c}-\hat{d}).
\end{equation}

The dissipative dynamics described in Eq.~(\ref{eq:TwoModeMEq}) stabilizes the state $\ket{\psi_{\rm ss}}$ defined in Eq.~(\ref{eq:TwoModeSS}).  Transforming this state to the ``localized" $\hat{a}$,\ $\hat{b}$ mode basis is equivalent to passing the state through a beam splitter.  It is well known that two-mode entangled coherent states can be generated by sending single-mode cat states into the arms of a $50/50$ beam-splitter \cite{Asboth2004}.  We thus find that in the localized basis
(and for $\mu_0 \ll \mu_1$), our steady state has the form of a two-mode entangled cat state:
\begin{equation}
	\label{eq_desiredEntangledCat}
	\ket{\psi_{\mathrm{ss}}} \approx \frac{1}{\sqrt{2}}
		\left[\ket{\mathcal{C}_{+}(\balpha)}_{a} \ket{\mathcal{C_{+}}(\balpha)}_{b} -\ket{\mathcal{C}_{-}(\balpha)}_{a} \ket{\mathcal{C_{-}}(\balpha)}_{b}\right].
\end{equation}

To determine the feasibility of our scheme, we need to express the Hamiltonian $\hat{H}_1$ in terms of the localized mode operators $\hat{a},\ \hat{b}$.  Before doing this, we first modify $\hat{H}_1$ to make it more symmetric, but without changing the steady state:
\begin{align}
	\hat{H}_2 & = \hat{H}_1 +
		\left(\Lambda \mu_1 \hat{c}^\dagger
		\left( \hat{c}^\dagger \hat{c} \right) \hat{d} + h.c. \right) \nonumber \\
	& =
		\Lambda \hat{c}^{\dagger}\left[\mu_0 \hat{d} + \mu_1 (\hat{c}^{\dagger}\hat{c}+\hat{d}^{\dagger}\hat{d})\hat{d}
			+ \nu \hat{d}^{\dagger} \right]  + h.c. \label{eq:H2}
\end{align}
We again consider the master equation
\begin{equation}
	\frac{d}{dt} \hrho =-i[\hat{H_2},\hrho]+\gamma\mathcal{L}[\hat{c}]\hrho. \label{eq:H2ME}
\end{equation}
It is easy to see that the state $\ket{\psi_{\rm ss}}$ (c.f.~Eq.~(\ref{eq:TwoModeSS})) continues to be the steady state of this master equation; this follows from the fact that the term we added to the Hamiltonian also annihilates this state.

We now consider the form of the nonlinear Hamiltonian $\hat{H}_2$ in the localized mode basis.  Defining for convenience
\begin{align}
	\Delta = \Lambda \mu_0, \>\> \lambda = \Lambda \nu / 2, \>\> U = \Lambda \mu_1,
	\label{eq:HParams}
\end{align}
we obtain
\begin{align}
 	 \frac{d}{dt} \hrho = -i[\hat{H}_2,\hrho]+\gamma \mathcal{L}[\hat{a} + \hat{b}]\hrho,
	 \label{eq:FinalMEq}
\end{align}
with
\begin{align}
	\hat{H}_2
	&=
	\Delta \hat{a}^{\dagger}\hat{a} + \lambda(\hat{a}^{\dagger}\hat{a}^{\dagger} + h.c.) + U \hat{a}^{\dagger}\hat{a}^{\dagger}\hat{a}\hat{a}
		\nonumber \\
		&- \Delta \hat{b}^{\dagger}
	\hat{b} - \lambda (\hat{b}^{\dagger}\hat{b}^{\dagger} + h.c.) -U \hat{b}^{\dagger}\hat{b}^{\dagger}\hat{b}\hat{b}.
	\label{eq:FinalH}
\end{align}

We see that the required Hamiltonian has a remarkably simple form, corresponding to the schematic in Fig.~\ref{fig_schematic}.
It describes two {\it uncoupled} nonlinear driven cavities whose Hamiltonians differ only by a sign.  Each
cavity has a Kerr nonlinearity (strength $U, -U$) and is subject to a parametric drive (drive detunings $\pm \Delta$, drive amplitudes $\pm \lambda$).
The only interaction between the two cavities comes from the non-local (but linear) dissipator in the master equation, arising from a linear interaction with a common reservoir.
Such dissipators have been discussed extensively and even realized experimentally (e.g.~\cite{Fang2017}) in many different contexts.  For the simple case where each cavity has the same resonance frequency, it can be generated by coupling both cavities to a transmission line with an appropriately chosen distance between the cavities \cite{Chang2012,Metelmann2015}.  Further details about implementation issues (including how to realize a repulsive Kerr interaction in circuit QED) are discussed in
Appendix \ref{app:Implementation}.

Eqs.~(\ref{eq:FinalMEq})-(\ref{eq:FinalH}) thus completely describe our dissipative scheme for the preparation and stabilization of entangled non-Gaussian states.
Note that the emergence of cat states and their adiabatic preparation in a single parametrically-driven Kerr cavity is the subject of several recent works \cite{Goto2016,Goto:2016aa,Ciuti2016,Puri2017}; our scheme leverages similar resources in two cavities (along with engineered dissipation) to now {\it stabilize} an entangled cat state.
Note that the amplitude $\balpha$ of our cat state (c.f.~Eq.~(\ref{eq:alpha})) in terms of the energy scales in the Hamiltonian is now $\balpha = i \sqrt{\lambda / U}$.

\section{Exact solution for the steady state}
\label{sec_exactSolution}

We now explore in more detail our analytic expression for the pure steady state of our system.  For simplicity, we work in the
non-local $\hat{c},\ \hat{d}$ mode basis.  In this basis the steady state $\ket{\psi_{\rm ss}}$ has the $\hat{c}$ mode in vacuum, while the $\hat{d}$ mode has a definite even photon number parity (c.f.~Eq.~(\ref{eq:OneModeSS})).   By solving a simple recursion relation (see Appendix \ref{app:Exact}), the even-parity state of the $\hat{d}$ mode is given by:
\begin{equation}
		 \Ket{\tilde{C}[\mu_0,\mu_1,\nu]}_d  =
		 \mathcal{N}  \sum_{n=0}^{\infty} \tilde{c}_{2n} \ket{2n}_d,
		 \label{eq:ExactSS}
\end{equation}
\begin{equation}
\label{eq_exactSteadyState}
	\tilde{c}_{2n} =
		\left(\frac{-\nu}{2 \mu_1}\right)^{n}\sqrt{\frac{(2n-1)!!}{(2n)!!}}
		\frac{\Gamma\left(\frac{1}{2}+\frac{\mu_0}{2 \mu_1} \right)}{\Gamma\left(n+\frac{1}{2}+\frac{\mu_0}{2\mu_1}\right)}.
\end{equation}
Here $\ket{2n}_d$ denotes a Fock state with $2n$ photons, $\Gamma(z)$ is the Gamma function, and $\mathcal{N}$ is an overall normalization constant.
Our ability to easily find an analytic form relies on the symmetry of the problem and the existence of a pure steady state; it is much simpler than more general methods based on the positive-P function \cite{Drummond1980,Ciuti2016,Hafezi2016}. Our solution method is reminiscent of that used in Ref.~\cite{Stannigel2012}, which studied dissipative entanglement in cascaded quantum systems; in contrast to that work, our approach does not require cascaded interactions, or even any breaking of time-reversal symmetry.

\begin{figure}[t]
\centering
\includegraphics[width=1\linewidth]{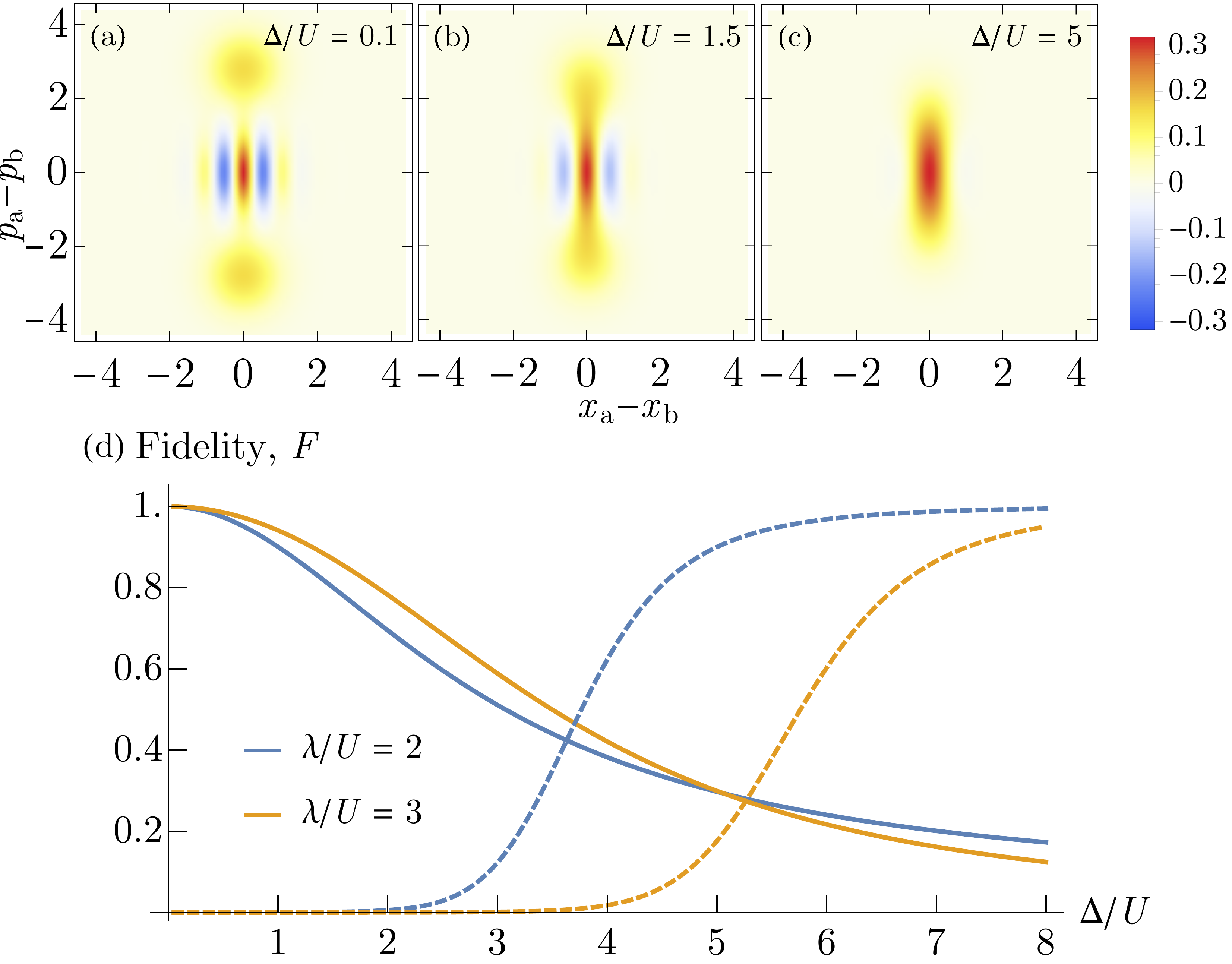}
\caption{(a-c) Wigner functions of the steady state (c.f.~Eq.~(\ref{eq:TwoModeSS})) for the non-local mode $\hat{d}$ at snapshots of $\Delta/U$. This is equivalent to a cut through the two-mode Wigner function for the local modes, defined by the combined quadratures $(\hat{x}_{a}-\hat{x}_{b})/\sqrt{2}$ and $(\hat{p}_{a} - \hat{p}_{b})/\sqrt{2}$, where $\hat{x}_{k}$ and $\hat{p}_{k}$ are the usual quadrature operators for mode $k$. As $\Delta/U$ increases, the evolution from an even-parity cat state to a squeezed state is clearly visible. (d) Fidelity of the exact steady state (c.f.~Eq.~\eqref{eq:ExactSS}) with both the squeezed state (dashed lines) obtained for $\Delta/U \rightarrow \infty$, and the even cat-state (solid lines) obtained for $\Delta/U \rightarrow 0$ (c.f.~Eq.~\eqref{eq_desiredEntangledCat}). The fidelity is defined by $F\left(\ket{\psi_1},\ket{\psi_2}\right) = \left|\bra{\psi_1} \psi_2 \rangle\right|^2$.
\label{fig_squeezedToCat}}
\end{figure}

The analytic expression for $\ket{\psi_{\rm ss}}$ allows us to confirm the expected parameter dependence of the state.  For $\mu_0 \gg \mu_1$, nonlinearity plays a minor role, and the state tends to a Gaussian squeezed state.  In contrast, for $\mu_0 \ll \mu_1$, the nonlinearity is crucial to the existence of a steady state, and $\ket{\psi_{\rm ss}}$ tends to the desired even cat-state.
Fig.~\ref{fig_squeezedToCat} uses the exact expression of the steady state $\ket{\tilde{C}}_d$ to show this evolution as a function of the ratio $\Delta / U = \mu_0 / \mu_1$.
The upper panel depicts the Wigner function for the $\hat{d}$ mode at three specific values of $\Delta/U$.  As described in the figure caption, this describes a cut through the two-mode Wigner function describing the state in the localized $\hat{a},\ \hat{b}$ basis.  One clearly sees the cross-over from an even cat-state to a squeezed state as $\Delta$ is increased.  This figure also demonstrates that even for modest values of $\Delta / U$, the fidelity with the desired even-parity cat state can be extremely high.

\section{Effect of imperfections}

While our ideal system always possesses an entangled pure steady state, imperfections in any realistic implementation will cause deviations from the desired dynamics.
The most important imperfection will come in the form of unwanted internal losses in each cavity.  To assess the impact of such losses, we add single-photon loss (at a rate $\kappa$ in each cavity) to our master equation:
\begin{equation}
	\label{eq_lossyMasterEquation}
	\frac{d}{dt} \hrho = -i[\hat{H}_2, \hrho] + \gamma \mathcal{L}[\hat{a} + \hat{b}]\rho + \kappa \mathcal{L}[\hat{a}]\hrho + \kappa \mathcal{L}[\hat{b}]\hrho.
\end{equation}

With the inclusion of single particle loss, the steady state will no longer be pure, and we cannot find an analytic closed form solution.  We instead numerically simulate the master equation.  At a heuristic level, the sensitivity to weak internal loss will depend on the typical timescale required by the ideal system to reach its steady state.  As discussed in detail in Appendix \ref{app:Rates}, this timescale can be extremely long, scaling as $e^{4 |\balpha|^2}$.  As a result, low internal loss levels (relative to $U$, $\lambda$, and $\Delta$) are required for good performance.

\begin{figure}[h]
\centering
\includegraphics[width=1\linewidth]{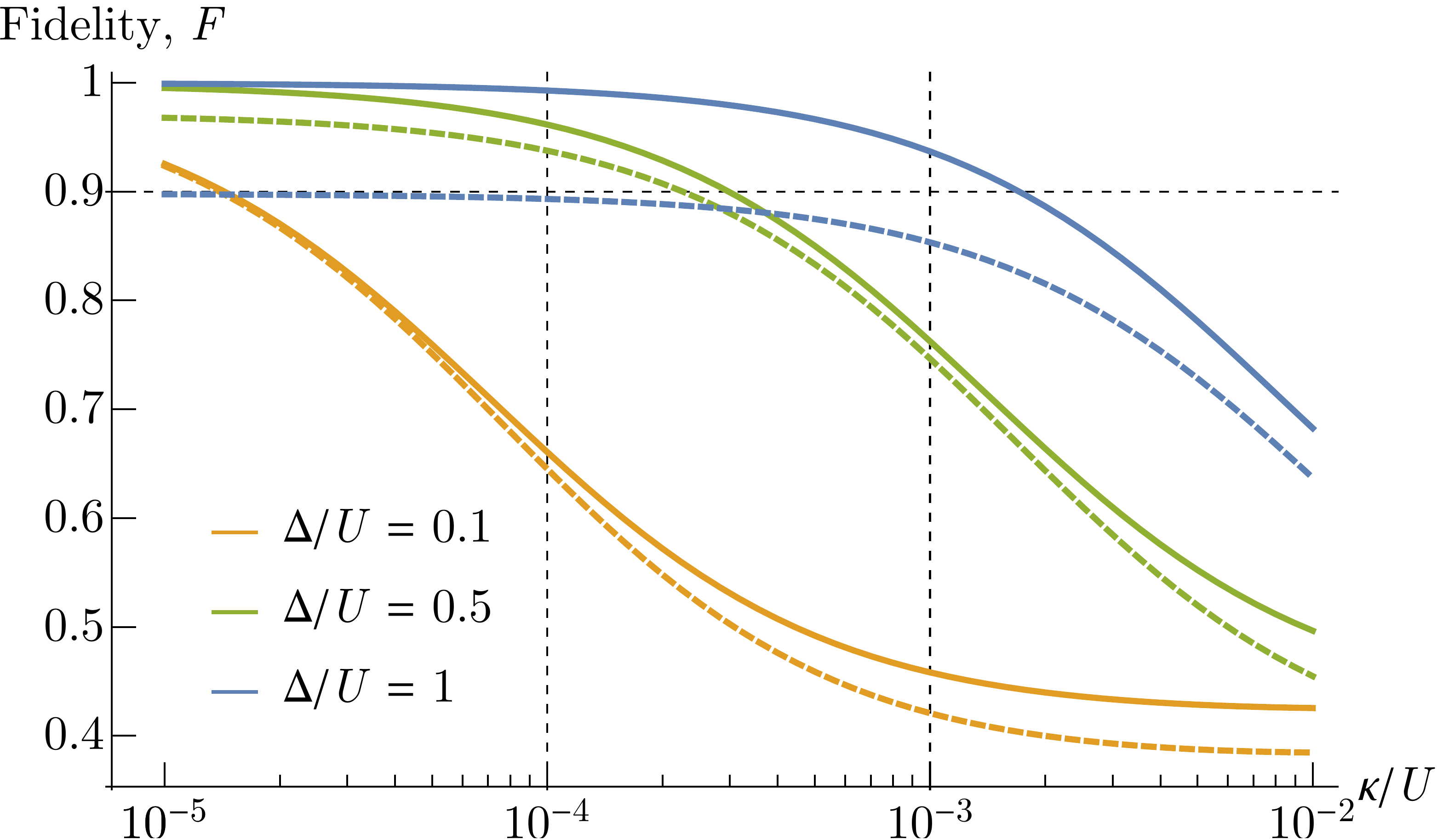}
\caption{Effect of intrinsic cavity losses $\kappa$ (identical for both cavities, c.f.~Eq.~\eqref{eq_lossyMasterEquation}) on the steady state, for various values of the
parametric drive detuning parameter $\Delta$.  Solid lines are the fidelity of the numerically-obtained steady state including loss, with the ideal loss-free steady state of Eq.~\eqref{eq:ExactSS}.  Dashed lines are the fidelity with the two-mode entangled cat state of Eq.~\eqref{eq_desiredEntangledCat} (which coincides with the ideal steady state when $\Delta/U \rightarrow 0$). For all curves the parametric drive strength is set to $\lambda/U = 2$ (thus setting the amplitude $\balpha = i \sqrt{2}$),  and the coupling rate $\gamma$ to the correlated dissipation is taken to be $\gamma = \sqrt{2} \Delta$ to maximize the speed of internal system dynamics (see Appendix \ref{app:Rates}). The fidelity is defined by $F(\rho_1,\rho_2) = \left|\text{Tr}\left[\sqrt{\sqrt{\rho_1} \rho_2\sqrt{\rho_2}}\right]\right|^2$.}
\label{fig_intrinsicLossFidelity}
\end{figure}

Fig.~\ref{fig_intrinsicLossFidelity} plots the fidelity of the numerically-obtained steady state with both the ideal, loss-free steady state $\ket{\psi_{\rm ss}}$ (c.f.~Eq.~(\ref{eq:TwoModeSS})), as well as with the ideal two-mode entangled cat-state of Eq.~\eqref{eq_desiredEntangledCat}, as a function of the internal loss rate $\kappa$.
We find that for $\Delta/U = 1$, we get $F \approx 0.95$ with the exact solution and $F \approx 0.86$ with the ideal cat state for a loss rate  $\kappa = 10^{-3} U$. This value of loss is compatible with state-of-the-art experiments in circuit QED.  An attractive Kerr nonlinearity in a 3D superconducting microwave cavity at the level of $U \sim$ 100-1000 kHz has been
realized \cite{kirchmair2012kerr}, while in state-of-the-art 3D cavities internal leakage rates are on the order of 10-100 Hz~\cite{reagor2013cavityLifetime, reagor2015cavityLifetime}.

Note that for the results in Fig.~(\ref{fig_intrinsicLossFidelity}), we have taken the coupling $\gamma$ to the engineered dissipation to be $\gamma = \sqrt{2} \Delta$.  As discussed in more detail in Appendix \ref{app:Rates}, this choice minimizes the timescale required to prepare the steady state, and thus helps minimize the effects of internal loss.
However, the fidelity is not extremely sensitive to this precise tuning of the value of $\gamma$.

A qualitative description of the effect of internal loss can be found in Appendix \ref{app:WigNeg}, where we show the Wigner functions of an example steady state with and without internal loss. Internal loss causes a reduction in the magnitude and area of Wigner function negativity, indicating a reduction in quantum correlations between the two modes. For realistic internal loss rates this reduction is minimal, and the state remains entangled.

Finally, while the ideal system requires a matching of the Hamiltonian parameters of the two cavities, the desired behaviour is still found for small mismatches.  For the same parameter choices as in Fig.~\ref{fig_intrinsicLossFidelity}, a drive or nonlinearity strength mismatch by 5\% or less still produces a steady state with $F > 0.95$. Further details can be found in Appendix \ref{app:Mismatch}.

\section{Conclusion}

We have a presented a relatively simple scheme that uses a modified version of a driven-dissipative Bose Hubbard dimer to stabilize a two-cavity entangled cat state.  The scheme uses detuned parametric drives applied to each cavity, and couples them only through a linear dissipative tunneling interaction that is mediated by an external reservoir.  While we have focused on the case where there is a unique steady state, by setting the parametric drive detunings $\Delta = 0$, one would instead stabilize a two dimensional manifold spanned by the entangled cat states (the states $\ket{2}$ and $\ket{1}$ defined in Eq.~(\ref{eqs:ECatsBasis})).

\section{Acknowledgements}

We acknowledge R.~P.~Tiwari and J.~M.~Torres for fruitful discussions. This work was supported by NSERC, and by the AFOSR MURI FA9550-15-1-0029.

\section*{APPENDICES}

\appendix

\section{Analytic expression for the steady state}
\label{app:Exact}
We derive here the exact steady state of the system described by the Hamiltonian $\hat{H}_2$ in Eq.~(\ref{eq:H2}) and the master equation in Eq.~(\ref{eq:H2ME}).
Motivated by the fact that only the $\hat{c}$ mode experiences any damping, we make an ansatz that there exists a pure-state steady state in which the $\hat{c}$ mode is in vacuum, i.e.
\begin{equation}
	\ket{\psi} = \ket{0}_{c} \otimes \sum_{n=0}^{\infty} \alpha_n \ket{n}_{d}.
\end{equation}
Here, $\alpha_n$ are arbitrary normalized coefficients and $\ket{n}_d$ are Fock states.

The above state is trivially a dark-state of the jump operator $\hat{c}$ appearing in the dissipator of the master equation in Eq.~(\ref{eq:H2ME}).  To be a steady state, it must then necessarily also be an eigenstate of the Hamiltonian.
Acting on $\ket{\psi}$ with $\hat{H}_2$ yields,
\begin{align}
	\nonumber
	\hat{H}_2 \ket{\psi} = \ket{1}_{c} \otimes \sum_{n = 0}^{\infty} &\alpha_n \big(\left[\Delta + U(n-1) \right]\sqrt{n} \ket{n-1}_{d} \\
	& + 2\lambda \sqrt{n+1}\ket{n+1}_{d}\big).\label{eq_HActsOnPsi}
\end{align}
where we have expressed $\mu_0,\mu_1$ and $\nu$ in terms of $\Delta, U$ and $\lambda$ using Eq.~(\ref{eq:HParams}).

The RHS of Eq.~(\ref{eq_HActsOnPsi}) has the $\hat{c}$ mode in the state $\ket{1}$, while in $\ket{\psi}$ it is in the state $\ket{0}$.  Thus, the only way that $\ket{\psi}$ can be an eigenstate of $\hat{H}_2$ is if it has zero energy.  We thus need $\hat{H}_2$ to annihilate $\ket{\psi}$, and therefore require that the coefficient of each $\ket{n}_d$ state in Eq.~(\ref{eq_HActsOnPsi}) vanish.  This leads to the recursion relation
\begin{equation}
	\left( \Delta + U n \right) \sqrt{n+1} \alpha_{n+1} =
		\begin{cases}
     0 & n=0, \\
       -2\lambda  \sqrt{n} \alpha_{n-1} & n \geq 1.
\end{cases}
	\label{eq:recursion}
\end{equation}

The above recursion relation does not mix the even and odd-parity subspaces.  Note that for $\Delta \neq 0$, the $n=0$ case of Eq.~(\ref{eq_exactSteadyState}) forces $\alpha_1 = 0$, and as a result, forces {\it all} odd $\alpha_j$ to be zero.  The solution thus has only even-parity Fock states.  Solving Eq.~(\ref{eq:recursion}) for the remaining even parity coefficients directly gives the solution presented in Eq.~(\ref{eq_exactSteadyState}), with $\alpha_{2n} \equiv \tilde{c}_n$.

Note also that if $\Delta = 0$, then the $n=0$ case of Eq.~(\ref{eq_exactSteadyState}) is trivially satisfied regardless of the value of $\alpha_{1}$, and one can find both even and odd parity steady states.  Thus, as discussed in the main text, a non-zero detuning $\Delta$ is crucial in order to obtain a unique steady state.

While the above discussion has identified a pure steady state for our system when $\Delta \neq 0$, the question remains whether this is a unique steady state.  In the case where $U=0$, the system is linear and one can exactly solve the system.  The linear system is stable for $\lambda \leq \Delta$, and in this regime the unique steady state is exactly that given by Eq.~(\ref{eq_exactSteadyState}).  For non-zero $U$, we have no rigorous analytic argument precluding the existence of multiple steady states (i.e.~in addition to the state we describe analytically here).  We have however investigated the model numerically over a wide range of parameters, and find that as long as $\Delta$ is non-zero, the steady state described by Eq.~(\ref{eq:ExactSS}) is indeed the unique steady state.

\section{Dynamical timescales}
\label{app:Rates}

We consider in this appendix the parameter dependence of the slow rates of the master equation in Eq.~(\ref{eq:FinalMEq}) that determine the typical time needed to prepare our non-Gaussian entangled steady state.  As discussed, this timescale sets the sensitivity to unwanted perturbations such as internal loss.  Note first that when the drive detuning $\Delta = 0$, each cavity has exact zero-energy eigenstates $\ket{\mathcal{C}_{\pm}(\balpha)}$ \cite{Goto2016,Goto:2016aa,Puri2017}.  For the interesting limit of a large cat state amplitude $\balpha$, the slowest timescales in our system correspond to dynamics within the four dimensional subspace spanned by these cat states.  Entangled cat states
form an orthonormal basis for this subspace:
\begin{align}
	\ket{\tilde{2}} &= \frac{1}{\sqrt{2}}\left(\ket{\mathcal{C}_+(\balpha)} \ket{\mathcal{C}_+(\balpha)} -
		\ket{\mathcal{C}_-(\balpha)} \ket{\mathcal{C}_-(\balpha)}\right), \nonumber \\
	\ket{\tilde{4}} &= \frac{1}{\sqrt{2}}\left(\ket{\mathcal{C}_+(\balpha)} \ket{\mathcal{C}_+(\balpha)} +
		 \ket{\mathcal{C}_-(\balpha)} \ket{\mathcal{C}_-(\balpha)}\right),\nonumber \\
	\ket{1} &= \frac{1}{\sqrt{2}}\left(\ket{\mathcal{C}_+(\balpha)} \ket{\mathcal{C}_-(\balpha)} -
		\ket{\mathcal{C}_-(\balpha)} \ket{\mathcal{C}_+(\balpha)}\right),\nonumber \\
	\ket{3} &= \frac{1}{\sqrt{2}}\left(\ket{\mathcal{C}_+(\balpha)} \ket{\mathcal{C}_-(\balpha)} +
		\ket{\mathcal{C}_-(\balpha)} \ket{\mathcal{C}_+(\balpha)}\right). \label{eqs:ECatsBasis}
\end{align}
The states $\ket{\tilde{2}},\ket{\tilde{4}}$ have even total photon number parity, while the states $\ket{1},\ket{3}$ have an odd total photon number parity.

It is useful to make a slight rotation of the even parity states:
\begin{align}
	\ket{2} & \equiv \cos \theta \ket{\tilde{2}} + \sin \theta \ket{\tilde{4}}, \\
	\ket{4} & \equiv -\sin \theta \ket{\tilde{2}} + \cos \theta \ket{\tilde{4}}.
\end{align}
where for large $|\balpha|$ we have
\begin{equation}
	\sin \theta \simeq 2 e^{-2|\balpha|^2}.
\end{equation}
Note that in the large $|\balpha|$ limit, $\ket{2} \rightarrow \ket{\tilde{2}}$, which is our expected entangled-cat steady state.

\begin{figure}[t]
\centering
\includegraphics[width=0.8\linewidth]{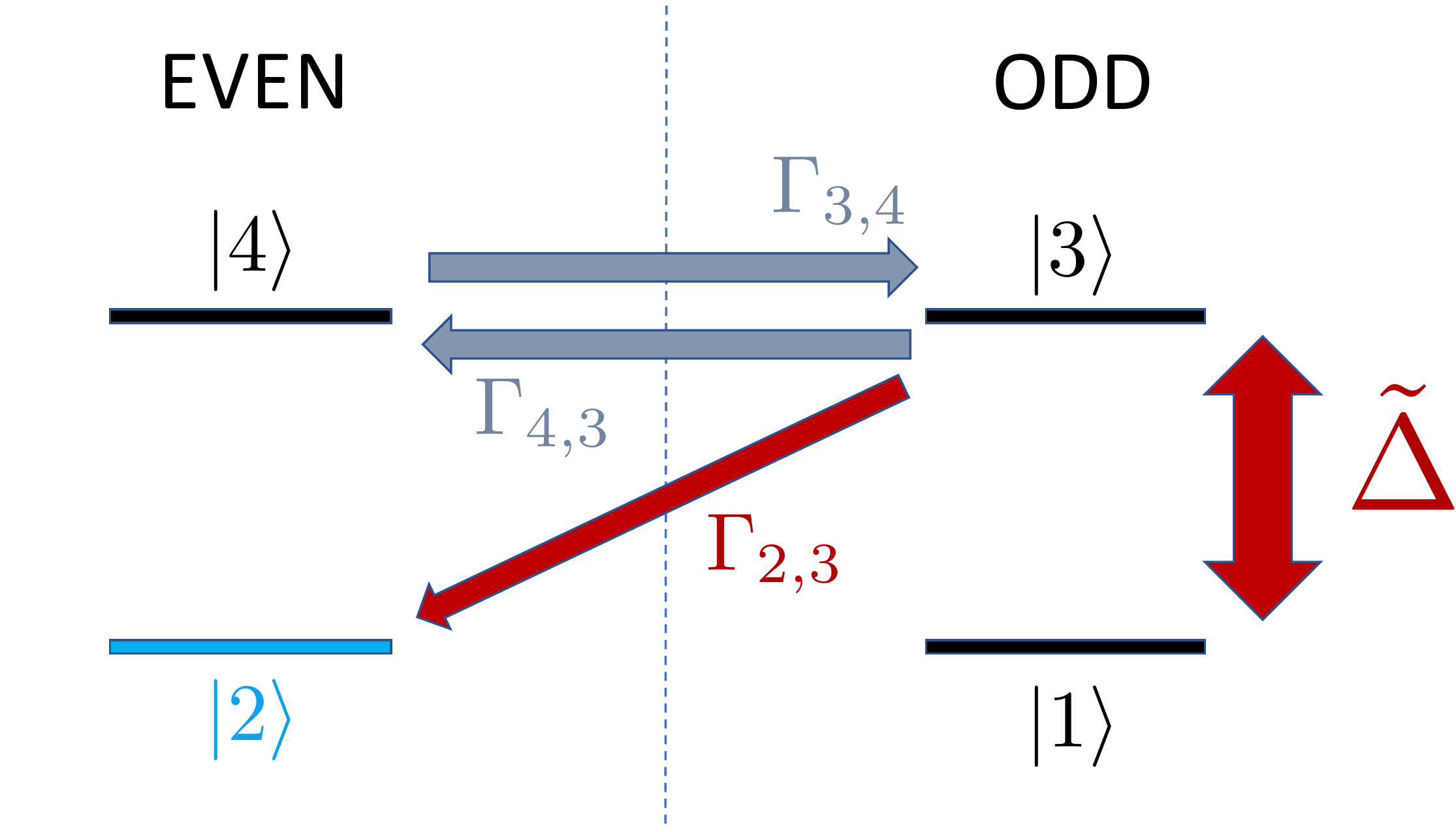}
\caption{Schematic showing the effective slow dynamics in the four dimensional manifold spanned by even and odd cat-states in each cavity.  There are two dark states of the dissipator in this subspace, denoted $\ket{2}$ and $\ket{1}$, with $\ket{2}$ the eventual steady state.  The detuning $\Delta$ of the parametric drives induces tunneling between the odd parity states $\ket{1}$ and $\ket{3}$, while dissipation induces the indicated transitions between states of different parity.  Exponentially small rates are indicates in dark red.    }
\label{fig:Rates}
\end{figure}

The states $\ket{1},\ket{2},\ket{3},\ket{4}$ will form a useful basis for our cat state subspace.  One finds that there are two distinct dark states in this subspace:  they correspond to the basis states $\ket{1}$ and $\ket{2}$.  If we now restrict our master equation to this cat-state subspace, we find the effective dynamics shown schematically in Fig.~\ref{fig:Rates}.  The detuning term in $\hat{H}_2$ becomes a coherent tunneling term of amplitude $\tilde{\Delta}$ between the odd parity states $\ket{1}$ and $\ket{3}$, whereas the dissipator gives rise to the three depicted incoherent transitions (with rates $\Gamma_{3,4}, \Gamma_{4,3}, \Gamma_{2,3}$).  One sees clearly from this schematic that, as expected, the dark state $\ket{2}$ will be our eventual steady state.

For large $|\balpha|$, one finds that $\tilde{\Delta}$ and $\Gamma_{2,3}$ become exponentially small:
\begin{align}
	\tilde{\Delta} & \simeq -4 |\balpha|^2 e^{-2 | \balpha |^2} \Delta, \\
	\Gamma_{2,3} & \simeq 16 | \balpha|^2 e^{-4 | \balpha |^2} \gamma.
\end{align}
The exponential suppression of these rates corresponds to the extremely weak overlap of the coherent states $\ket{\balpha}$ and $\ket{-\balpha}$.
The above processes will then form the bottleneck in having populations relax to the state $\ket{2}$, and will determine the slow timescales characterizing state preparation.

The superoperator describing this reduced master equation can be readily diagonalized, yielding the relevant dynamical rates.  In the large $\balpha$ limit, we find that the rate-limiting slow relaxation between the dark states $\ket{1}$ and $\ket{2}$ is described by
\begin{equation}
	\Gamma_{\rm rel} =
		\frac{1}{2} \left(
			2 \Gamma_\Delta + \Gamma_{2,3} - \sqrt{\left(2 \Gamma_\Delta\right)^2 + \left(\Gamma_{2,3}\right)^2}
			\right),
\end{equation}
where we have defined the effective incoherent tunneling rate between the state $\ket{1}$ and the $\{\ket{3}$, $\ket{4}\}$ bright-state manifold as:
\begin{equation}
	\Gamma_{\Delta}
	= \frac{4 \tilde{\Delta}^2}{\Gamma_{2,3} + \Gamma_{4,3}} \simeq 16 | \balpha |^2 e^{-4 | \balpha |^2} \frac{\Delta^2}{\gamma}.
\end{equation}
The rate $\Gamma_{\rm rel}$ describes an effective two step process where population first incoherently tunnels from $\ket{1}$ to the bright state manifold (rate $\Gamma_{\Delta}$), and then relaxes to the steady state $\ket{2}$ (rate $\Gamma_{2,3}$).  Note that $\Gamma_{\rm rel} \neq \left( \Gamma_{\Delta}^{-1} + \Gamma_{2,3}^{-1} \right)^{-1}$, as the incoherent tunneling process described by $\Gamma_{\Delta}$ can also move population the wrong way, i.e.~from $\ket{3}$ back to the state $\ket{1}$.

It is instructive to consider the behaviour of $\Gamma_{\rm rel}$ for small and large $\Delta / \gamma$:
\begin{equation}
	\Gamma_{\rm rel}
	\simeq
		\begin{cases}
     	 		8 | \balpha |^2 e^{-4 | \balpha |^2} \gamma
				& \text{ if } \Delta \gg \gamma, \\
      			16 | \balpha |^2 e^{-4 | \balpha |^2} \frac{\Delta^2}{\gamma}
				& \text{if } \Delta \ll \gamma.
\end{cases}
\end{equation}
As expected, the relaxation rate vanishes in both these extreme limits, as the relaxation process from $\ket{1}$ to $\ket{2}$ involves both the $\tilde{\Delta} \propto \Delta$ tunneling process and the dissipative processes $\Gamma_{2,3} \propto \gamma$.  One finds that for fixed $\Delta$ and $\balpha$, $\Gamma_{\rm rel}$ is maximized when the damping rate $\gamma$ is chosen to be $\gamma = \sqrt{2} \Delta$.  This motivates the choice used in Fig.~\ref{fig_intrinsicLossFidelity}.

\section{Implementation issues}
\label{app:Implementation}

\subsection{Parametric drives}

In the lab frame, the Hamiltonian for two parametrically driven cavities is
\begin{align}
	\nonumber\hat{H}_{\rm Lab} &=
	\omega_a \hat{a}^{\dagger}\hat{a}  + \lambda(\hat{a}^{\dagger}\hat{a}^{\dagger}e^{-i\omega_{p_a}t} + h.c.) \\
	&+ \omega_b \hat{b}^{\dagger}\hat{b}- \lambda (\hat{b}^{\dagger}\hat{b}^{\dagger}e^{-i\omega_{p_b}t} + h.c.),
\end{align}
where $\omega_{a/b}$ are the cavity resonance frequencies and $\omega_{p_{a/b}}$ are the frequencies of the parametric drives on each cavity. To obtain the quadratic part of the Hamiltonian in Eq.~\eqref{eq:FinalH}, we go to a frame rotating at $\omega_{p_a}/2$ for cavity $\hat{a}$ and $\omega_{p_b}/2$ for cavity $\hat{b}$, and choose the parametric drive frequencies such that $\Delta = \omega_a - \omega_{p_a}/2 = \omega_{p_b}/2 - \omega_b$.

It is most convenient to consider a situation where the parametric drives are at the same frequency, i.e.~$\omega_{p_a} = \omega_{p_b} \equiv \omega_p$, as this simplifies phase locking of the two drives, which is required to maintain the relative sign difference in the two drive amplitudes. In this case, the cavities must have different frequencies, and the parametric drive frequency is simply taken to be the average cavity frequency:  $2\omega_p = \omega_a + \omega_b $.  This results in the detuning parameter $\Delta$ being given by $\Delta = (\omega_a - \omega_b)/2$.

Choosing identical pump frequencies also has a strong advantage when considering the correlated dissipator needed for our scheme, as now, this dissipator can be implemented in a completely passive manner.  One can for example simply couple the two cavities to a waveguide, as discussed in the next subsection.  For this choice of pump frequencies, we can go to an interaction picture at $\omega_p/2$ for the waveguide coupled to the cavities, with the result that the cavity-waveguide interaction is time independent for both cavities.  As discussed in the next subsection, one can then derive the needed $\hat{a} + \hat{b}$ dissipator by arranging the distance between the cavities appropriately.

If instead the cavities are not pumped at the same parametric drive frequency, the simple waveguide scheme for generating the needed dissipator fails, as we can no longer define a global interaction frame where both cavity-waveguide interactions are time-independent.  In this case, a correlated dissipator can still be achieved using frequency conversion via engineered dissipation with active elements (see e.g.~\cite{Sliwa:2015aa,Lecocq:2017aa,Lecocq:2016aa}).

\subsection{Non-local dissipator}

As discussed in the main text, the non-local dissipator in Eq.~(\ref{eq:FinalMEq}) (corresponding to correlated single photon loss) can be generated by coupling the two cavities to a one-dimensional transmission line or waveguide.  A full derivation of this is provided in Ref.~\cite{Chang2012} and in Appendix B of Ref.~\cite{Metelmann2015}.  In general, coupling two cavities to a waveguide and then eliminating the waveguide will generate not only a dissipator of the form in Eq.~(\ref{eq:FinalMEq}), but also a direct, Hamiltonian tunneling interaction between the cavities.  The relative strength of these two kinds of couplings is a function of the distance $d$ between the attachment points of the cavities to the waveguide.  For $d = m  \tilde\lambda $ (where $m$ is an integer, and $\tilde\lambda$ is the wavelength of a waveguide excitation at frequency $\omega_p/2$), the induced Hamiltonian tunneling interaction vanishes, and the only effect is the desired non-local dissipator with the correct phase.  Note that this result requires that the relevant dissipative dynamics is slow compared to the propagation time $ d / v_g$, where $v_g$ is the waveguide group velocity.  The slow dynamics of our system (see Appendix \ref{app:Rates}) means that this condition can easily be satisfied.

\subsection{Positive $U$}

While the use of Josephson junctions in superconducting circuits to induce effective attractive Kerr nonlinearities is by now routine (see e.g.~\cite{kirchmair2012kerr}), it is also possible to generate the repulsive interaction required in our scheme.  One option is to use a qubit based on an inductively shunted Josephson junction (such as a flux qubit or fluxonium \cite{Manucharyan113}).  The effective Hamiltonian describing such a system has the form \cite{GirvinHouches2014}
\begin{equation}
	\hat{H}_{\rm qb} =
		E_C \hat{n}^2 + \frac{1}{2} E_L \hat{\phi}^2 - E_J \cos\left(\hat{\phi} + \phi_g \right),
\end{equation}
where $E_C$ is the capacitive charging energy, $E_L$ is the inductive energy associated with the shunt inductance, and $E_J$ is the Josephson energy.  $\hat{\phi}$ ($\hat{n}$) is the junction phase (charge) operator.  The parameter $\phi_g$ describes a flux bias of the system, and is proportional to the externally applied magnetic flux. If one applies half a flux quantum, $\phi_g = \pi$, then one effectively changes the sign of the Josephson potential.  Expanding the cosine potential in this case yields a quartic term with a positive coefficient, corresponding to a repulsive Kerr interaction.  If one now weakly and non-resonantly couples a cavity to this qubit circuit (i.e.~the dispersive regime), the cavity will inherit this repulsive nonlinearity.

An alternate approach would be to use a standard transmon qubit \cite{Koch:2007aa}, but work in the so-called straddling regime, where the cavity frequency is between the frequencies for the ground to first excited state and first excited to second excited state transitions of the transmon. In the dispersive regime, one can treat the interaction between the transmon and the cavity perturbatively, and standard time-independent perturbation theory to fourth order \cite{Zhu:2013aa} yields an effective self-Kerr interaction for the cavity given by
\begin{align}
	\hat{H}_{\rm Kerr} = U_{\rm eff}\hat{a}^\dagger\hat{a}^\dagger\hat{a}\hat{a},\ U_{\rm eff} = \frac{g^4U_{\rm qb}}{2\delta^3(\delta + \frac{U_{\rm qb}}{2})}.
\end{align}
Here $\delta = \omega_{01} - \omega_{c}$ is the detuning, and $g$ the coupling between the transmon's ground to first exited state transition and the cavity, and $U_{\rm qb}$ is the anharmonicity of the transmon. For a transmon $U_{\rm qb} \approx -E_C$, i.e. $U_{\rm qb}<0$. If $\delta < 0$, or $\delta >0$ and $\delta + U_{\rm qb}/2 >0$, then $U_{\rm eff} <0$ and the cavity nonlinearity is attractive. However, if $\delta > 0$ and $\delta + U_{\rm qb}/2 < 0$, which lies in the straddling regime (defined by $\delta + U_{\rm qb} < 0$), then $U_{\rm eff} > 0$ and the cavity will inherit a repulsive nonlinearity from the transmon.

Other, more recent circuit designs also allow for strong repulsive Kerr nonlinearities, and even the ability to tune the strength of the nonlinearity \emph{in situ} \cite{Frattini:2017aa,Zhang:2017aa}.

\section{Wigner Function Negativity}
\label{app:WigNeg}

While the fidelity is a good quantitative measure of the success of our scheme in the presence of internal loss, it would also be interesting to understand how the quantum correlations in the steady state diminish as a result of internal loss. To that end, we compare the Wigner functions of the steady state with no internal loss, and that with finite internal loss rate, where the lossy steady state (panel (b)) has $90\%$ fidelity with the ideal steady state.

The relevant information is contained in the negativity of the Wigner function, and while the negativity is reduced by internal loss, even for a $10\%$ fidelity loss there is still significant negativity. Further, the interference fringes of the Wigner function of the non-local mode $\hat{d}$ correspond to the entanglement between the local modes $\hat{a}$ and $\hat{b}$. As can be seen, the fringes are still clearly visible in the lossy case, indicating that entanglement between the local modes remains.

\begin{figure}[h!]
	\includegraphics[width=\columnwidth]{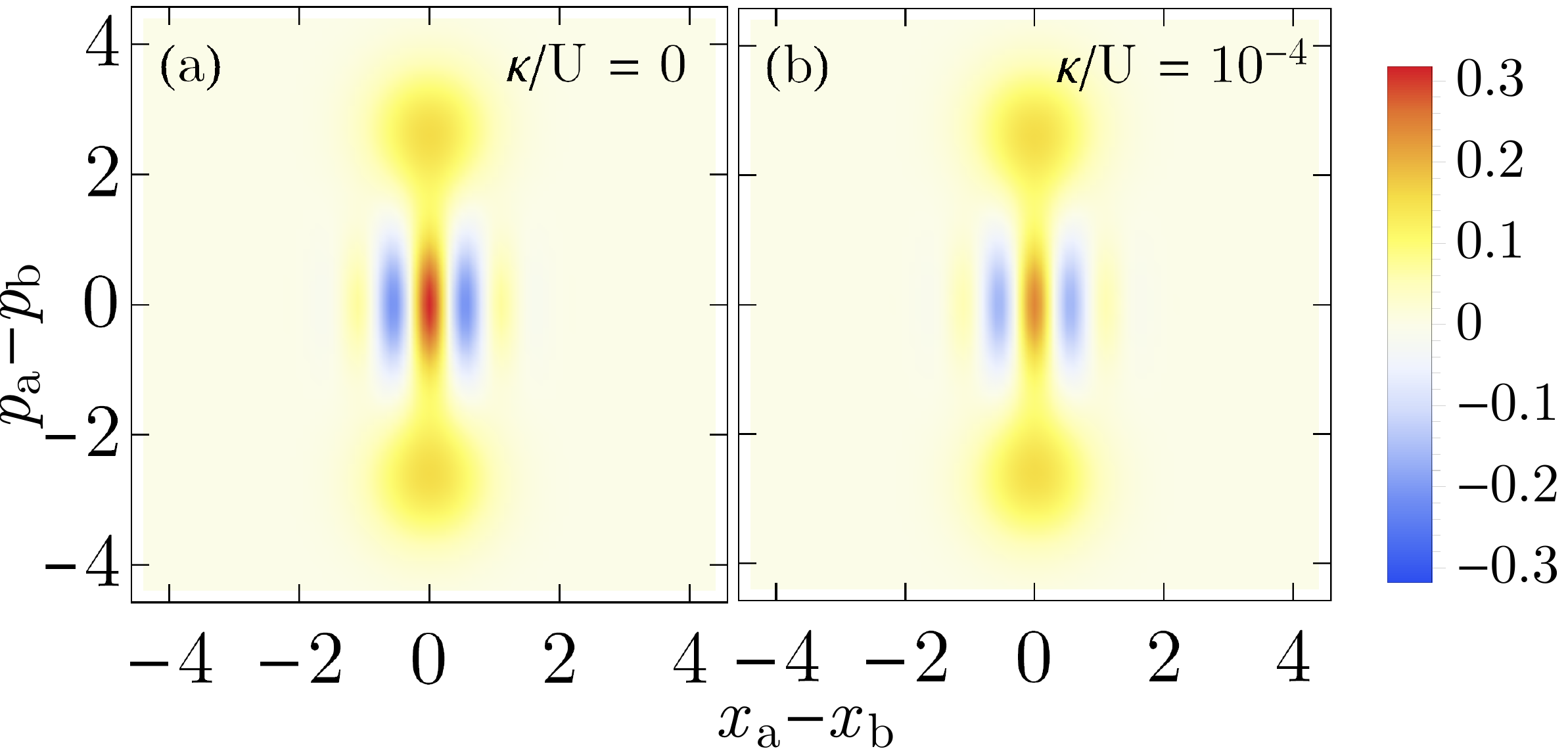}
	\caption{Wigner function of the steady state of the non-local mode $\hat{d}$ for (a) zero internal loss (c.f.~Eq.~(\ref{eq:TwoModeSS})) and (b) finite internal loss $\kappa/U =2\times10^{-4}$ (identical for both cavities, c.f.~Eq.~\eqref{eq_lossyMasterEquation}). We use system parameters of $\lambda/U=2$, $\Delta/U=0.5$, $\gamma/U=1/\sqrt{2}$, which for (b) corresponds to a fidelity of $F\approx0.9$, as seen in Fig.~\ref{fig_intrinsicLossFidelity}. The minimum Wigner function negativity is $W=-0.21$ for zero internal loss in panel (a), and $W=-0.16$ for finite internal loss in panel (b).}
	\label{fig_WigNeg}
\end{figure}

\section{Parameter mismatch}
\label{app:Mismatch}

\begin{figure}[!h]
\centering
\includegraphics[width=1\linewidth]{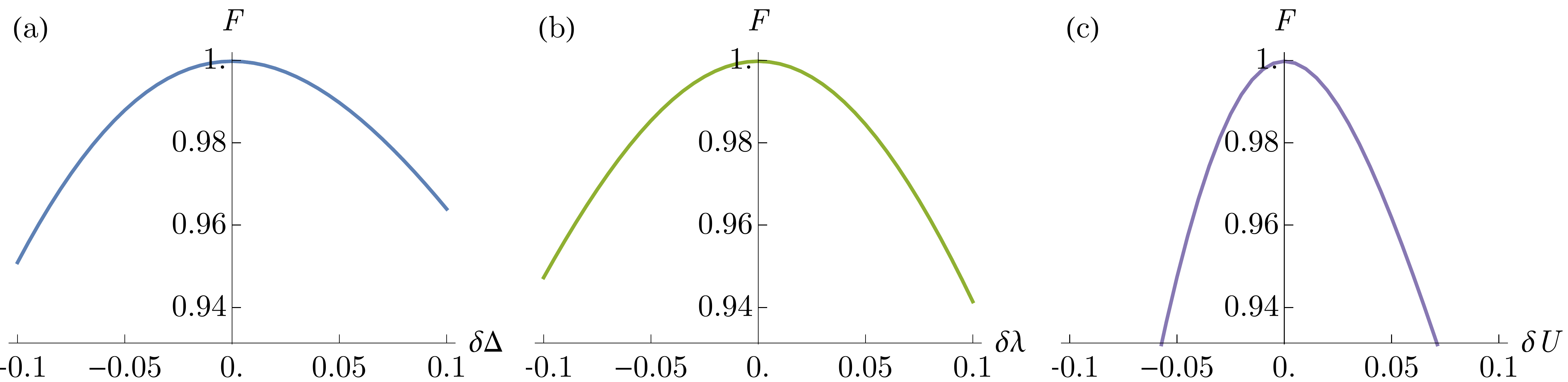}
\caption{Fidelity $F$ for our system with mismatched (a) cavity detuning $\Delta$, (b) parametric drive $\lambda$ and (c) self-Kerr nonlinearity $U$, as described by the modified Hamiltonian in Eq.~\eqref{eq_mismatchHamiltonian}. We compare to the exact steady state $\ket{0}_c\otimes\Ket{\tilde{C}}_d$ of Eq.~\eqref{eq:ExactSS}. The mismatch parameters $\delta \Delta$, $\delta \lambda$, $\delta U$ are dimensionless relative differences between the two cavities. For all curves the parametric drive strength $\lambda/U = 2$, the detuning $\Delta/U = 1$, and the correlated dissipation rate $\gamma/U  =2$. As can be seen, $F > 0.95$ can be obtained for up to $5\%$ relative mismatch.}
\label{fig_parameterMismatching}
\end{figure}

The pure steady state of our system requires matching of parameters between two Kerr cavities. In this appendix, we quantify the loss of fidelity from having the second cavity imperfectly matched with the first. To do so, we numerically calculate the steady state of the Hamiltonian
\begin{equation}
\begin{aligned}
\label{eq_mismatchHamiltonian}
\hat{H}_{\mathrm{mis}} &=\Delta \left[\hat{a}^{\dagger}\hat{a} - (1+\delta \Delta)\hat{b}^{\dagger}\hat{b}\right]\\ &+\lambda\left[\hat{a}^{\dagger}\hat{a}^{\dagger}-(1+\delta\lambda)\hat{b}^{\dagger}\hat{b}^{\dagger}+h.c.\right]\\ &+U\left[\hat{a}^{\dagger}\hat{a}^{\dagger}\hat{a}\hat{a}-(1+\delta U)\hat{b}^{\dagger}\hat{b}^{\dagger}\hat{b}\hat{b}\right].
\end{aligned}
\end{equation}
where $\delta\Delta$, $\delta \lambda$, and $\delta U$ are the dimensionless mismatch in the cavity detuning, parametric drive strength, and self-Kerr nonlinearity respectively. The original system is recovered for $\delta\Delta = \delta \lambda = \delta U=0$. Figure~\ref{fig_parameterMismatching} shows the fidelity of the numerically-computed steady state with the ideal steady state $\ket{0}_c\otimes\Ket{\tilde{C}}_d$ of Eq.~\eqref{eq:ExactSS}. We find that $F > 0.95$ is maintained for mismatches as large as $5\%$. Having imperfectly matched cavities thus poses no significant concern for the scheme.

\bibliographystyle{apsrev4-1}
\bibliography{DriveDispBHDBibliography}

\end{document}